\documentclass[rapids]{jfm}

\usepackage{graphicx}
\usepackage{newtxtext}
\usepackage{newtxmath}
\usepackage{natbib}
\usepackage{hyperref}
\hypersetup{
    colorlinks = true,
    urlcolor   = blue,
    citecolor  = blue,
    linkcolor  = blue
}
\usepackage{layouts}

\newcommand\Ray{\mbox{\textit{Ra}}}  	
\newcommand\COTwo{CO$_{2}$}  	

\newcommand{\cs}[1]{{{\hypersetup{linkcolor=blue,citecolor=blue}{\color{blue} #1}}}}

\shorttitle{Optimal displacement ventilation}
\shortauthor{R.\,Yang and others}
\title{Optimal ventilation rate for effective displacement ventilation}
\author{Rui Yang$^1$, Chong Shen Ng$^1$\thanks{c.s.ng@utwente.nl}, Kai Leong Chong$^1$\thanks{k.l.chong@utwente.nl}, Roberto Verzicco$^{1,2,3}$ and Detlef Lohse$^{1,4}$\thanks{d.lohse@utwente.nl}}

\affiliation{$^1$Physics of Fluids Group, Max Planck Center for Complex Fluid Dynamics, J.\,M.\,Burgers Center for Fluid Dynamics and MESA+ Research Institute, Department of Science and Technology, University of Twente, 7500AE Enschede, The Netherlands\\[\affilskip]
$^2$Gran Sasso Science Institute - Viale F. Crispi, 7 67100 L'Aquila, Italy\\[\affilskip]
$^3$Dipartimento di Ingegneria Industriale, University of Rome `Tor Vergata', Roma 00133, Italy\\[\affilskip]
$^4$Max Planck Institute for Dynamics and Self-Organisation, 37077 G{\"o}ttingen, Germany
}

\begin{document}

\maketitle

\begin{abstract}
Indoor ventilation is essential for a healthy and comfortable living environment. 
A key issue is to discharge anthropogenic air contamination such as CO$_2$ gas or, more seriously, airborne respiratory droplets. Here, by employing direct numerical simulations, we study the mechanical displacement ventilation with the realistic range of air changes per hour (ACH) from $1$ to $10$. For this ventilation scheme, a cool lower zone is established beneath the warm upper zone with the interface height $h$ depending on ACH. For weak ventilation, we find the scalings relation of the interface height $h\sim$ ACH$^{3/5}$, as suggested by Hunt \& Linden (Build.\,Environ., vol.\,34, 1999, pp.\,707-720). Also, the \COTwo~concentration decreases with ACH within this regime. However, for too strong ventilation, the interface height $h$ becomes insensitive to ACH, and the CO$_2$ concentration remains unchanged. Our results are in contrast to the general belief that stronger flow is more helpful to remove contaminants. We work out the physical mechanism governing the transition between the low ACH and the high ACH regimes. It is determined by the relative strength of the kinetic energy from the inflow, potential energy from the stably-stratified layers, and energy loss due to drag. Our findings provide a physics-based guideline to optimize displacement ventilation.
\end{abstract}


\section{Introduction}
Ventilation is important to optimize the air quality in indoor spaces \citep{linden1999fluid,awbi2003ventilation,morawska2020a}. There are many reasons for achieving effective ventilation, such as thermal comfort, energy savings and to minimize the pollutants such as \COTwo~or aerosols contributing to the spread of infectious diseases \citep{fisk2000health,li2007role,who2020ventilation,cdc2021ventilation}. Recently, indoor ventilation has become increasingly important considering that the majority of COVID-19 infections are caused by a- or pre-symptomatic infected individuals, and that, by speaking and breathing, tiny salive droplets are released into the environment, accumulate \citep{abkarian2020,pohlker2021respiratory} and persist in the air for long \citep{bourouiba2020turbulent,bourouiba2021fluid,chong2021extended}. To measure the amount of respiratory contaminants in the room, one often relies on the fact that tiny droplets, aerosols, or droplet nuclei (i.e. dried out droplet) behave like tracers and are simply carried along by the air flow. Therefore they evolve similarly to another tracer produced through exhalation, namely carbon dioxide (\COTwo). Since \COTwo~is easily measured, it is often regarded as surrogate for indoor airborne contaminants \citep{vonPohle1992carbon,rudnick2003risk}.

There are two basic ventilation types \citep{linden1999fluid,awbi2003ventilation,chenvidyakarn2013buoyancy,bhagat2020effects}. First, for mixing ventilation, air is circulated throughout the room, leading to a uniform distribution of temperature and contaminants \citep{linden1999fluid}, and the purpose of this ventilation type is usually thermal comfort. The second type is displacement ventilation which has the benefit of displacing contaminants to the upper level of the room where it is extracted, so that occupants breathe fresher air \citep{bhagat2020displacement}. In this case, the air inlet is at the bottom of the room and the air outlet is diagonal to it or at least semi-diagonal at the top. This arrangement results in a flow through the room interior from bottom to top, with the exhaled droplets or aerosols being discharged upwards. The flow in the interior is stratified: Each occupant produces an upward thermal plume (the typical heat output of a person at rest is 80 W) due to the body heat \citep{craven2006computational,bhagat2020effects}, which leads to a density stratification with the warm, lighter air layer above the cooler, heavier counterpart (see for example figure \ref{fig1}$a$). Respiratory contaminants exhaled together with a warm vapour puff that are not entrained by the main body plume tend to accumulate between the stratified air layers \citep{bolster2007contaminants,bhagat2020effects} to form the so-called lock-up layer \citep{qian2006dispersion,zhou2017lock}. 

Early studies of displacement ventilation were mainly focused on experiments and theories. \cite{sandberg1990stratified} were the first to discuss displacement ventilation with a source of buoyancy, which is referred to as natural displacement ventilation since there is no mechanical extraction. \cite{linden1990emptying} further investigated this phenomenon in an enclosure with inflow at the bottom and outflow at the top. They demonstrated the presence of thermal stratification with two layers of uniform temperatures separated by a horizontal interface (see e.g.\,figure \ref{fig:FlowSetup}). They also derived how the layer height $h$ of the lower cold zone depends on the effective area of the vent. The stratification is also affected by the number of occupants. With separately distributed buoyancy sources, multiple layers can form, leading to a more gradual change of temperature in the domain \citep{cooper1996natural,livermore2007natural}. 

When the displacement ventilation is driven by mechanical extraction or an input wind \citep{hunt1999fluid,hunt2001steady,hunt2005displacement}, the interface height $h$ is obtained from matching the total extraction rate $Q$ and the buoyancy sources, giving the hallmark formula:
\begin{equation}
Q=cn^{2/3}B^{1/3}(h-h_v)^{5/3}. \label{eqn:BuoyPlumeUnstratifiedBG}
\end{equation}
Equation \eqref{eqn:BuoyPlumeUnstratifiedBG} is obtained by scaling arguments of a self-similar buoyant plume in an unstratified background \citep{linden1990emptying,mtt1956,schmidt1941turbulent}. Here, $B$ is the buoyancy flux produced by $n$ occupants and $h_v$ the ``virtual origin'' of the thermal plume, i.e.\,the height at which the plume would start if it were a point buoyant plume. The empirical constant $c$ has an approximate value of $c\approx0.105$ \citep{mtt1956,linden1999fluid}.  Given that the formula is based on some simple scaling arguments and is derived using some assumptions, one wonders on how well the clean zone height $h$ prediction works for practical extraction rates and in what range it holds.

In this study, we use direct numerical simulation (DNS) to study the mechanical displacement ventilation with a wide and practically relevant range of air change units per hour (ACH). To reproduce realistic indoor air flows, the flow is fully coupled to the temperature, the \COTwo~and water vapour concentration fields. Previous works on indoor ventilation has mainly focused on the design for energy efficiency, whereas due to COVID-19 pandemic the recent objective is on how the ventilation can effectively remove the respiratory contaminant. 

Previous numerical studies mainly employed large-eddy simulation (LES) or turbulence models \citep{davidson1989ventilation,durrani2015evaluation,van2017accuracy} while numerical results obtained by DNS are scarce. However, LES and turbulence models perform differently in different airflow cases \citep{zhang2007evaluation}. Simulating flows in the near-wall region is more challenging than those in the free flow region \citep{piomelli2002wall} due to the dominance of smaller vortices. LES also becomes more questionable once temperature and \COTwo~are coupled since with LES the turbulent heat flux can be significantly underestimated \citep{taylor2008direct}. Thus the dataset from our \cs{DNS} can be seen as benchmark for comparison with results by LES or turbulence models, which are unavoidable for large parameter space studies.

The paper is organised as follows: In \S\,\ref{sec:FlowSetup}, the flow setup and the governing equations are introduced. In \S\,\ref{sec:Results}, we discuss how the lock-up effects change with increasing ACH (\S\,\ref{subsec:LockUpEffect}), explain the change in the layer formation by an energy balance (\S\,\ref{subsec:PEAndKE}), analyse globally and locally averaged concentration values (\S\,\ref{subsec:LocalStats}), and discuss the influence of different ventilation setups on the lock-up effect (\S\,\ref{subsec:DiffFlowConfig}). Finally, in \S\,\ref{sec:Conclusions}, we summarise our findings and provide an outlook for future work.

\section{Flow Setup}\label{sec:FlowSetup}
\begin{figure}
\centering
\centerline{\includegraphics[width=0.9\columnwidth]{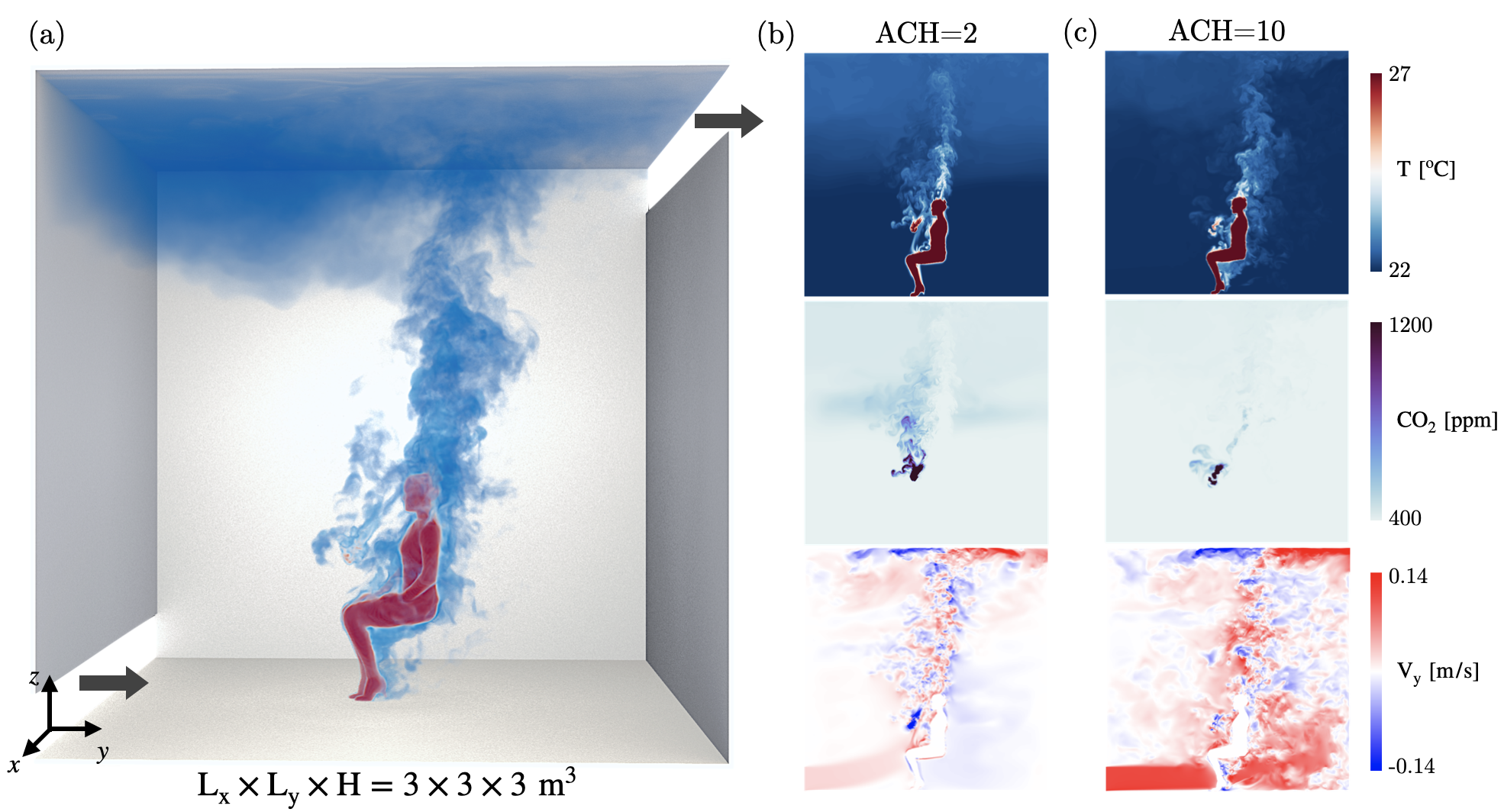}}
\caption{\label{fig:FlowSetup}($a$) Illustration of the simulation setup with the body plume, breathing flow, and arrows indicates the inlet and outlet flows and ($b$) Temperature, \COTwo~field and horizontal velocity of ventilated flows for ($b$) ACH~$=2$, where clean thermal stratification, the interface height $h$, and lock-up \COTwo~layer are observed, and ($c$) for ACH~$=10$ with much stronger turbulent activity behind the body.}
\label{fig1}
\end{figure}

We consider a room of dimensions $3\times3\times3$m$^3$, having an idealised lower-level inlet flow and higher-level outlet flow on opposite walls, as sketched in figure \ref{fig:FlowSetup}($a$). For simplicity, both inlet and outlet have the same heights (0.3m) and widths (3m). The ambient temperature, relative humidity (RH), and \COTwo~volume fraction of the room are considered fresh and homogeneous initially and set to values acceptable for indoor comfort, i.e.\,$\theta_{\textit{amb}}=22^\circ$C, 40\% and 0.04\%, respectively. Correspondingly, the inlet flow is set to be the same as the initial ambient conditions, whereas outflow boundary condition is imposed at the outlet. For the walls, no-slip, impermeable, and adiabatic boundary conditions are imposed.

A seated occupant within the room produces thermal buoyancy (from body temperature), high vapour concentration (from exhalation), and respiratory contaminants (represented by \COTwo). The flow motion is assumed to be incompressible, $\partial_i u_i=0$. To capture complete interactions of these anthropogenic sources, the flow is fully coupled with the heat, water vapour concentration, and \COTwo~concentration by applying the Boussinesq approximation. The governing equations read
\begin{subeqnarray}
\partial_t u_i + u_j \partial_j u_i &=& -\rho_{\textit{ref}}^{-1}\partial_i p + \nu\partial^2_j u_i + g \delta_{i3} (\beta_\theta \theta + \beta_v v - \beta_c c), \\
\partial_t \phi + u_j \partial_j \phi &=& \kappa_\phi\partial^2_j \phi. \label{eqn:GovEqn}
\end{subeqnarray}
\returnthesubequation
The scalar variable $\phi$ represents either temperature $T$, vapour mass fraction $\gamma$, or \COTwo~volume fraction $c$, all of which are simultaneously taken into consideration. We define $\rho_{\textit{ref}}$ as the reference density, $\beta_{\phi}$ as the isobaric thermal expansion coefficient of the fluid for each scalar, $\nu$ is the kinematic viscosity, and $\kappa_{\phi}$ the diffusivity for each scalar, for example, $\kappa_{T}$ the thermal diffusivity. All of these quantities are assumed to be independent of temperature. Accordingly, the gas Prandtl number is $\Pran \equiv \nu/\kappa_{T} = 0.71$.

The thermal Rayleigh number of the occupant, $\Ray_{\textit{ocpt}} \equiv g\beta_\theta \Delta h_{\textit{ocpt}}^3/(\nu \kappa_{T}) = 8.8\times 10^8$, where $\Delta \equiv \theta_{\textit{ocpt}}-\theta_{\textit{amb}}$ and $h_{\textit{ocpt}}=1.2$m. The grid resolutions are $768^3$ with a clipped Chebychev clustering in the $z$-direction to resolve gradients at the upper and lower walls. Our resolutions have been validated by performing grid refinement studies and by comparing with the body thermal plume velocity profile from experiments \citep{craven2006computational}, (see Supplementary Material). Equations (\ref{eqn:GovEqn}$a$,$b$) are solved by DNS using the multi-scalar second-order finite difference method with a fractional third-order Runge--Kutta scheme \citep{verzicco1996finite,ostilla2015multiple}. The body in the room is modelled by the immersed boundary method with no-slip boundary, fixed body temperature, and zero \COTwo~and vapour concentrations.

The interactions are expected to be non-trivial because of the competing effects of hot humid air, which is lighter than the ambient, with the heavier \COTwo. The occupant body temperature is set as $\theta_{\textit{ocpt}} =27^\circ$C \citep{houdas2013human}, which is the mean of the periphery and ambient temperature. The exhaled breath is set to 27$^\circ$C (given that the temperature drops from mouth at 37$^\circ$C), with RH~$=100$\% and \COTwo~volume fraction of 4\%. The tidal period and volume of the breath is set to 4s and 2l, respectively, matching experimental measurements \citep{gupta2010characterizing}. Since our focus is on the large-scale flows in the room, and also to make our simulations tractable, the breath is modelled as a Gaussian spatial source at a distance of $\approx 20$cm from the mouth and angled $60^\circ$ below horizontal. A visualisation of the flow field is available in the Supplementary Movie. (given in the supplementary material)

The main control parameter of the simulations is the ACH, where 1ACH~$=10$ liters per second per person following convention \citep{bhagat2020effects}. This control parameter determines the height $h$ of the lower clean zone according to \eqref{eqn:BuoyPlumeUnstratifiedBG}, which is achieved after some transient. In \S\,\ref{subsec:DiffFlowConfig}, we will examine the generality of our findings by varying the placement of the inlet and outlet.

\section{Results} \label{sec:Results}

\subsection{Influence of ACH on the lock-up effect} \label{subsec:LockUpEffect}
\begin{figure}
\centering
\centerline{\includegraphics[width=1.0\columnwidth]{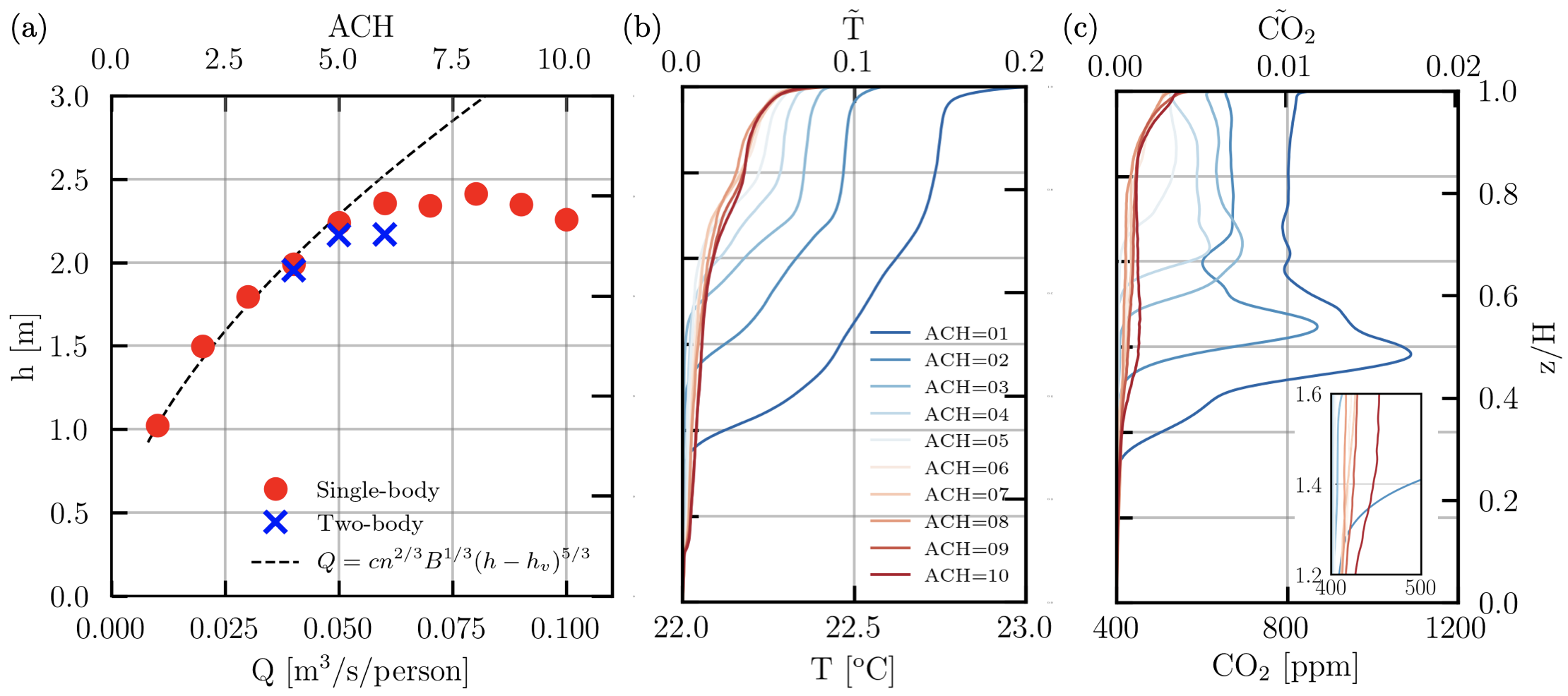}}
\caption{\label{fig:fig2}($a$) Plot of layer height $h$, based on the coordinate of the steepest temperature gradient between upper part of the inlet and lower part of the outlet, versus ACH. In ($a$), at ACH~$\lesssim 5$, the trend of $h$ is predicted by \eqref{eqn:BuoyPlumeUnstratifiedBG}, which is shown by the black dashed line with $h_v=0.24~m$. Mean profiles for ($b$) temperature and ($c$) \COTwo~concentration for various ACH-values. }
\end{figure}

We begin by analysing the clean-zone height, $h$, defined as the height below which the air maintains the same properties as at the inflow (see figure \ref{fig:FlowSetup}). Typically, clean zones are designed to be taller than occupants to ensure a contaminant-free occupied space. 

In presence of thermal sources, a stably stratified environment can exist within the room \citep{bolster2007contaminants} where a lower cooler air is separated by the upper warmer air by a temperature interface at height $h$. Contaminants such as aerosols and fomites that are entrained by the thermal sources can accumulate at this interface \citep{bolster2007contaminants}, and commonly referred to as the lock-up effect \citep{qian2006dispersion,zhou2017lock}. To illustrate this point, in figure \ref{fig:fig2}, $h$, the mean temperature profiles, and \COTwo~concentrations are plotted for various ACH.

In figure \ref{fig:fig2}($a$), we find that the interface height $h$ increases from ACH~$=1$ to $5$. This increasing trend is in remarkably good agreement with \eqref{eqn:BuoyPlumeUnstratifiedBG}, shown as dashed line in figure \ref{fig:fig2}($a$), indicating that the thermal body plume can be well approximated by assuming a simplified buoyant plume source in accordance with \cite{mtt1956}. However, when ACH $\gtrsim 5$, relation \eqref{eqn:BuoyPlumeUnstratifiedBG} no longer holds and $h$ remains roughly constant instead. This implies that the simplified assumptions cannot be extended to the largest ACH values. We will see later that this is because of the dominance of the inflow resulting in vastly different flow structures in the room. We call this large ACH regime the inflow-dominant regime. The weak influence of $h$ on ACH for high ACH values is also reflected in the temperature and \COTwo~profiles (figure \ref{fig:fig2}$b$,$c$), which remain roughly unchanged for ACH $\gtrsim 5$. To further test the robustness of these results, we conducted three additional simulations with two occupants in the room (shown as crosses in figure \ref{fig:fig2}$a$), which are also in agreement with the general trend. 

Using the \COTwo~concentration as an indicator of respiratory contaminants \citep{vonPohle1992carbon}, in figure \ref{fig:fig2}($c$), we find that the lock-up effect is evident for ACH $\lesssim 5$, as shown by the peaks in the \COTwo~concentration. At ACH~$\gtrsim 5$, the \COTwo~concentration is substantially reduced as compared to ACH~$=1$, suggesting that the contaminant layer depletes at large ACH and as expected for mechanical displacement ventilation \citep{bhagat2020displacement}. However, although the lock-up layer diminishes at large ACH, there is a modest increase in \COTwo~concentration close to head height at $z\approx 1.2$m (see figure \ref{fig:fig2}($c$) with inset). This implies that, at ACH~$\gtrsim 5$, contaminants become dispersed at lower heights which contributes to poorer air quality. The findings from this simplified setup highlight that careful design of indoor ventilation configuration is necessary to ensure optimal removal of contaminants. 


\subsection{Regime transition explained by balance of potential and kinetic energy} \label{subsec:PEAndKE}
\begin{figure}
\centering
\centerline{\includegraphics[width=1.0\columnwidth]{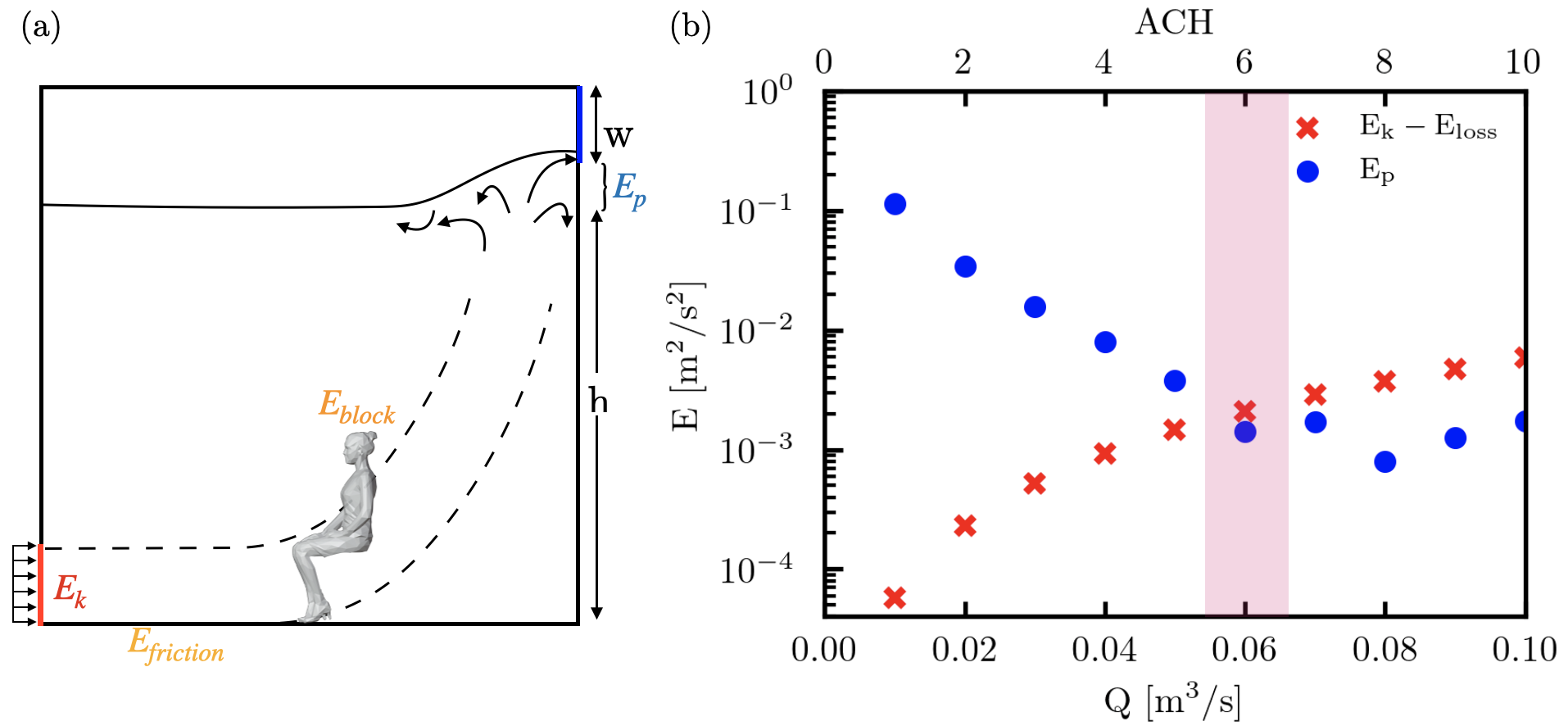}}
\caption{\label{fig3}($a$) Sketch of the potential energy, $E_p$, and kinetic energy, $E_k$, in the displacement ventilation flow. ($b$) Plot of $E_k$ and $E_p$ versus ACH. The transition to a stable $h_{\textit{layer}}$ in figure \ref{fig:fig2}($a$) correlates with $E_k \sim E_p$ (vertical \textcolor{blue}{shaded} area) when ACH~$\approx 6$, implying that the influence of mechanical mixing is diminished by a stronger inflow when ACH$\gtrsim 6$.}
\end{figure}

We next explain why there is transition to the inflow-dominant regime by examining the relative strength of the energies in the system. As depicted in figure \ref{fig3}($a$), three types of energies (per unit mass) can be identified: (i) Kinetic energy $E_k=(1/2)u^2$. (ii) Potential energy due to the stable stratification $E_p=N^2\Delta h_{\textit{layer}}^2$. (iii) Energy losses $E_{\textit{loss}}$ from friction and blockage. Here, $u$ is the inflow velocity, $N=\sqrt{\beta_\theta g \mathrm{d}T/\mathrm{d}z}$ the buoyancy frequency, where $\beta_\theta$ is the thermal expansion coefficient, $g$ the gravitational acceleration, $\mathrm{d}T/\mathrm{d}z$ the temperature gradient at the layer interface, $\Delta h_{\textit{layer}}$ the effective thickness of upper layer (with height of the outlet vent $w$ being subtracted) expressed as $\Delta h_{\textit{layer}}\equiv H-w-h$ where $h$ is the height of clean zone. Thus, the potential energy can be rewritten as $E_p=\beta_\theta g \mathrm{d}T/\mathrm{d}z (H-w-h)^2$. 

The energy loss in (iii) consists of two parts, which are the friction from the ground $E_{\textit{friction}}$ and the drag due to the blockage of the occupant $E_{\textit{block}}$. We estimate $E_{\textit{friction}}$ by $\frac{1}{2}C_f u^2$ with the friction coefficient for turbulent flow $C_f \equiv 2/H\int_0^{H/2} 0.664 \Rey_y^{-1/2}\,\mathrm{d}y \approx 0.03$--$0.01$ \citep[c.f.\,][]{pope2001turbulent}, where $\Rey_y(y) \equiv u y/\nu$. The other part of the energy loss is $E_{\textit{block}}$ which is expressed as $E_{\textit{block}}=\frac{\alpha}{2}C_d u^2$. Here, the geometrical factor $\alpha$ takes into account the partial blockage by the occupant (estimated by the multiple of cross section area of the occupant and the height of the occupant over the domain volume) and its value is approximately $0.06$. The drag coefficient $C_d \approx 0.7$ \citep{wang2012effects}.

In figure \ref{fig3}($b$), we compare the residual kinetic energy $E_k-E_{\textit{loss}}$ to the potential energy $E_p$ for various ACH. When ACH~$<5$, the inflow is still relatively weak compared to the strong thermal stratification established in the room and therefore, $E_p$ is dominant. Upon increasing ACH, the influence of the inflow becomes more dominant: At around ACH~$=6$, the residual kinetic energy becomes larger and overtakes $E_p$ (see inset of figure \ref{fig3}$b$). This energy argument explains the transition to the inflow-dominant regimes in figure \ref{fig:fig2}. As can be seen in figure \ref{fig:fig2}($c$), the layer height remains the same for large enough ACH. 
The physical explanation is that the strong inflow breaks the stratified layer near the outlet and directly leaves the room (see figure \ref{fig3}$a$). Therefore, the excessive kinetic energy no longer contributes to making the upper stratified layers thinner. Instead, a strong turbulent wake is generated behind the person, as shown for ACH~$=10$ in figure \ref{fig:FlowSetup}($c$).

\subsection{Globally and locally-averaged temperature and CO$_2$ concentration} \label{subsec:LocalStats}

\begin{figure}
\centering
\centerline{\includegraphics[width=1.0\columnwidth]{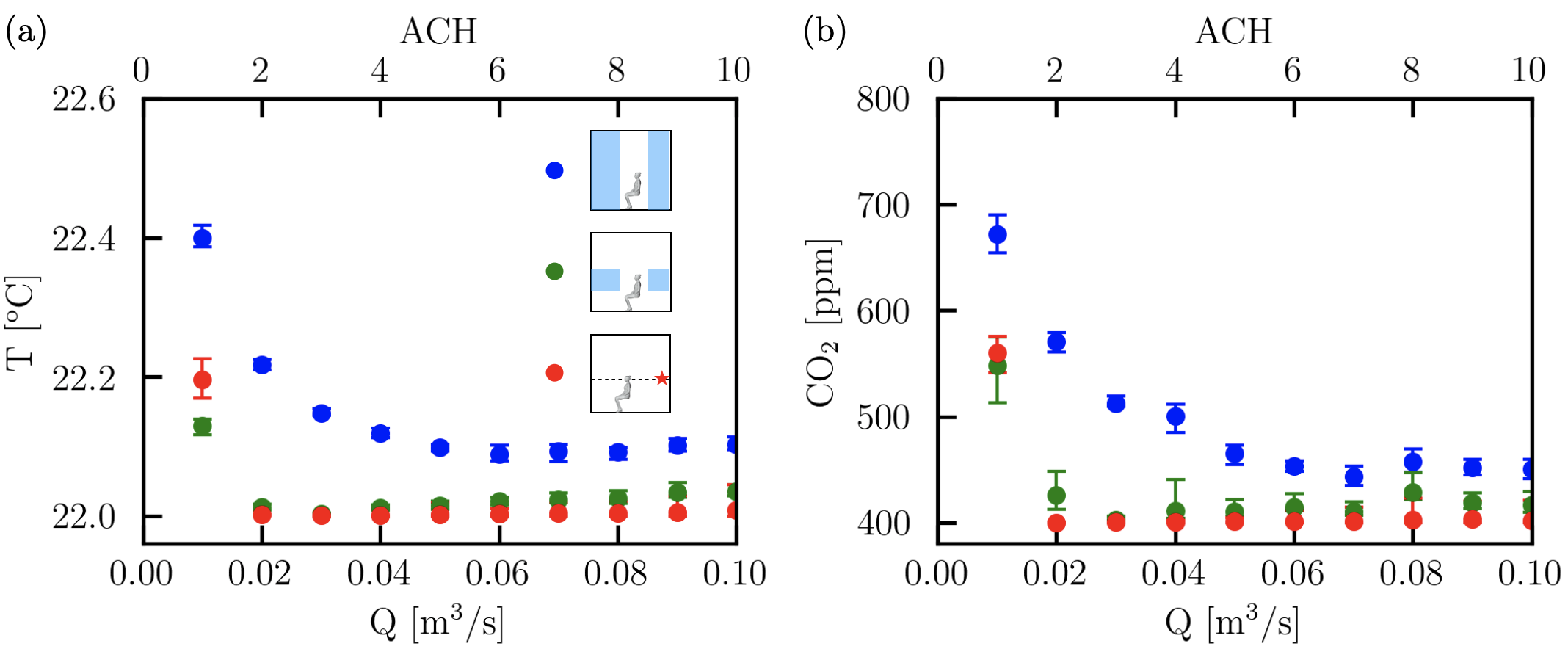}}
\caption{\label{fig:fig4}Plots of global average (without direct neighborhood of the person), local average at a height band $0.3\ m<h<1.5\ m$, and point measurement for ($a$) statistically stationary mean temperature and ($b$) mean \COTwo~concentration versus ACH, see sketches. With increasing ACH, both temperature and \COTwo~concentration are reduced to the ambient values.}
\end{figure}

While spatial profiles are useful for quantifying the interface height $h$ and the lock-up effect, it is also instructive to compute integral-averaged quantities since they represent measurable metrics for real-world cases. Therefore, in figure \ref{fig:fig4}, we plot the spatially-averaged temperature and \COTwo~concentrations versus ACH. The plane averaging is performed on the longitudinal plane and the region near the occupant is omitted. We consider three different ways of averaging, see the sketches in the inset of figure \ref{fig:fig4}($a$).

From figure \ref{fig:fig4}, in all these cases the averaged temperature and \COTwo~concentration eventually decrease to the ambient values with increasing ACH. In particular, local and point-wise measurements are reduced to ambient values when ACH\,$\gtrsim$\,2. In contrast, the global values are larger than the ambient values and plateaus at ACH\,$\gtrsim$\,5. These trends highlight two crucial points: Firstly, measurements at head-height provide an inadequate view of the temperature and \COTwo~concentration values \citep{mahyuddin2012review}. The inaccuracy is mainly caused by the non-trivial flow organisation, where a stably stratified background flow is separated by a temperature interface. Secondly, increasing ACH larger than 5 does not alter the global temperature value and \COTwo~concentration, which further supports the fact that an optimum ACH exists for mechanical displacement ventilation.

\subsection{Influence of different flow configurations} \label{subsec:DiffFlowConfig}
\begin{figure}
\centering
\centerline{\includegraphics[width=1\columnwidth]{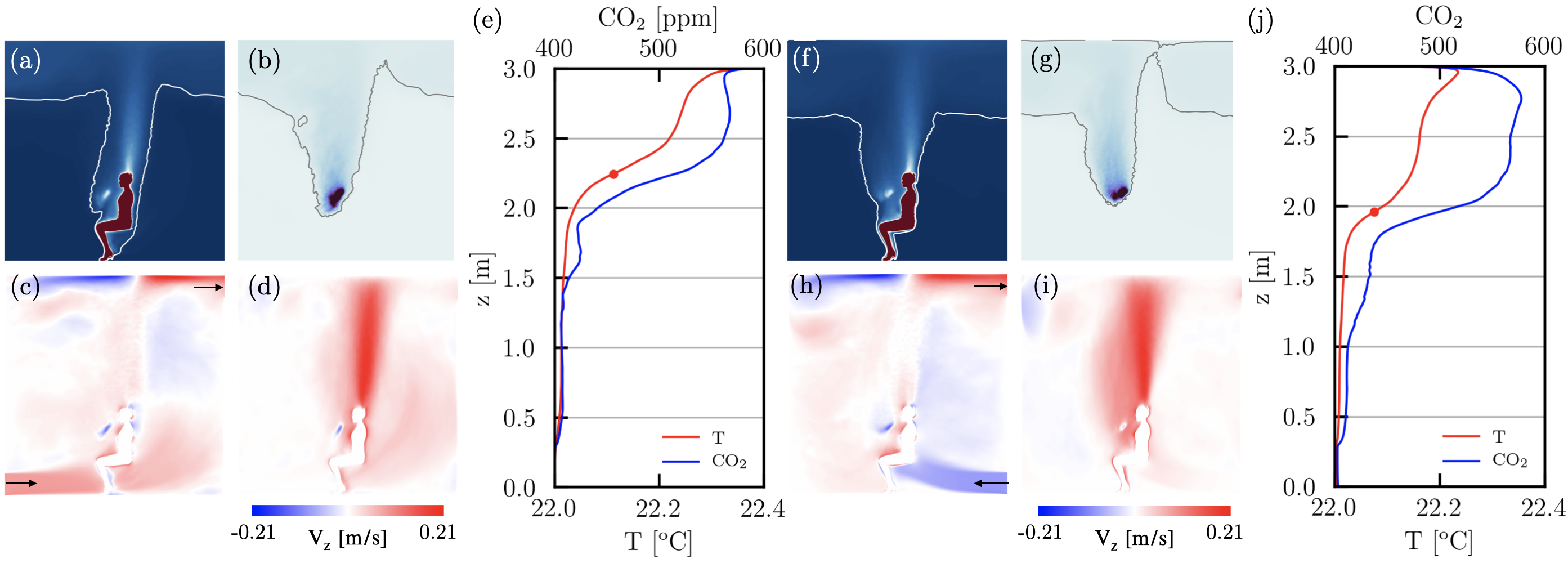}}
\caption{\label{fig:comparevent}Comparison of mean flow fields for two-sided ventilation ($a$-$e$) and one-sided ventilation ($f$-$j$). The 2D contours show: ($a$,$f$) temperature, ($b$,$g$) \COTwo~concentration, ($c$,$h$) horizontal velocity, and ($d$,$i$) vertical velocity, the colormap is kept the same as shown in figure \ref{fig:FlowSetup}. The vertical mean profiles of temperature and \COTwo~are shown in ($e$,$j$), where $h$ is denoted by the circle symbol.}
\end{figure}
To test the sensitivity to different inflow setups, we simulated additional cases with air inflow and outflow on the same side (see figures \ref{fig:comparevent}$f$-$j$). 
As seen from these figures, the stratified temperature field for this case and the original case (figure \ref{fig:comparevent}$a$-$e$) are quite similar (figure \ref{fig:comparevent}$a$,$f$).
Just as before, the region of high \COTwo~concentration is closely related to the inflection point in the mean temperature profile $T(z)$ (indicated for example by the isocontour of 500\,ppm in figure \ref{fig:comparevent}$b$,$g$).
On closer inspection, some key differences can be identified. For example, $h$ is much lower in the one-sided case ($\approx$1.9m, see circled marker in figure \ref{fig:comparevent}$j$) and as a result the \COTwo~concentration is also higher at that $z$-level. 
Some differences between the profiles are indeed expected since the large-scale flow structures are clearly different for the two-sided and one-sided ventilation cases. In particular, when we inspect the vertical velocity in the periphery above the body (see figure \ref{fig:comparevent}$i$), we find that there is a significant upward velocity contribution from the inflow. This upward contribution influences the \textit{effective} buoyant flux from the body, causing a stronger temperature mixing and thus lower interface height $h$.



\section{Conclusions and outlook} \label{sec:Conclusions}
In summary, we have shown and discussed a regime transition of mechanical displacement ventilation by studying a wide range of ACH. When the ventilation rate is low (ACH~$<5$), the observed flow structure agrees with the classical stratified flow structure of displacement ventilation, where there is a clean and cooler zone established beneath the warm contaminated zone. In this case, the height $h$ for the lower clean zone scales with ACH as $h\sim$~ACH$^{3/5}$, \cs{cf.\,\eqref{eqn:BuoyPlumeUnstratifiedBG}.} 
However, when the ventilation rate becomes high (ACH~$\gtrsim5$), we observe the inflow-dominant regime in which $h$ becomes insensitive to ACH: There is a major change in flow structure where the stratified layer partially breaks down. By measuring the mean \COTwo~concentration at the head-level, we find that the local concentration remains unchanged with increasing ACH in the inflow-dominant regime, implying that too large ACH-values do not help to remove more contaminants. Finally, we reveal that the transition ACH occurs when the kinetic energy of the inflow balances the sum of the potential energy of the stratified layer and the energy losses. To achieve effective displacement ventilation, an optimum ACH can be chosen based on this transition point, which may depend on the room geometry.


Motivated by mitigating the pandemic of COVID-19, many studies on indoor ventilation have been emerging recently \citep{morawska2020a,morawska2020time,bhagat2020effects,bhagat2020displacement}. Given that the parameter space is vast, it is computationally demanding to obtain a generalized view on indoor ventilation. Nevertheless,
by examining the energetic balance in the room with simple formulas, one can gain insight into the optimum ACH range for different scenarios, such as cases with different sized rooms, number of occupants, etc. LES and other turbulent models are more suitable for exploring the wide parameter space. Our DNS results of ventilation flows can be used for validations of these simulation methods.
Lastly, other factors could also influence our results and merit further investigation, such as when occupants are moving between rooms \citep{mingotti2020mixing}, or when occupants are performing different respiratory activities \citep{pohlker2021respiratory}.

\section*{Acknowledgements}
This work was funded by the Netherlands Organisation for Health Research and Development (ZonMW), project number 10430012010022: ``Measuring, understanding \& reducing respiratory droplet spreading'', the ERC Advanced Grant DDD, Number 740479, and Foundation for Fundamental Research on Matter with Project No. 16DDS001, which is financially supported by the Netherlands Organisation for Scientific Research (NWO). We acknowledge PRACE for awarding us access to MareNostrum in Spain at the Barcelona Computing Center (BSC) under the project 
2020235589, SURFsara (a subsidiary of SURF cooperation, the collaborative ICT organization for Dutch education and research), and Irene at Tr\`{e}s Grand Centre de calcul du CEA (TGCC) under PRACE project 2019215098.

The authors report no conflict of interest.


\end{document}